# Speed Control Security System For safety of Driver and Surroundings

Vishesh Vishal Ahire
Vishwakarma Institute of Technology,
Pune, Maharashtra, India
Vishesh.ahire24@vit.edu

Yash Badrinarayan Amle
Vishwakarma Institute of Technology,
Pune, Maharashtra, India
yash.amle24@vit.edu

Akshada Nanasaheb Waditke
Vishwakarma Institute of Technology,
Pune, Maharashtra, India
akshada.waditke24@vit.edu

Ojas Nitin Ahire
Vishwakarma Institute of Technology,
Pune, Maharashtra, India
ojas.ahire24@vit.edu

Amey Mahesh Warnekar
Vishwakarma Institute of Technology,
Pune, Maharashtra, India
amey.warnekar24@vit.edu

Ayush Ganesh Ahire
Vishwakarma Institute of Technology,
Pune, Maharashtra, India
ayush.ahire24@vit.edu

Prashant Anerao
Vishwakarma Institute of Technology,
Pune, Maharashtra, India
 prashant.anerao@vit.edu

*Abstract — the speed control security system is best suited for the task of slowing the speed of a vehicle during rash driving as the Driver is over speeding the circuit captures the images of the lanes witch decides the speed of the road the car is currently on this input is further provided to the ESP-32 micro Prosser module in the car switch compiles this data with the data received for the RPM sensor of the car and decides whether the car is over speeding or not in case of over speeding a signal is send by the ESP to the Arduino witch actuates the dc motor used in the car to reduce the speed of the car by the use of a hydraulic brake system actuated by a DC motor.*

## 1.Introduction:

Road traffic incidents happen every day across the globe as individuals injured and perish from over speeding. From overwhelming thresholds of injury and fatality for the average person and moving pedestrian, to the loss of property as well since these occur in places where intention for vehicles is to speed up and hit harder. Moreover, where speed limits, speed cameras, stop lights, manual signal directing fail relative to speed decreases - but not to the extent of ideal speed increases - there needs to be a speed control security system that at a minimum acknowledges over speeding and automatically works against it.

This innovative research project is a **Speed Control Security System (SCSS)**, an automated vehicle speed control security system that engages automatic speed reduction every time the user/pedestrian experiences over speeding beyond a pre-established, intended threshold. It involves a series of sensors, a microcontroller and a wireless, automated approach so that speed control engagement takes priority over actual engagement of acknowledgment of an over speeding in the moment.

Thus, automatic speed reduction not only protects everyone on the road, international speed accidents are decreased, protection from an individual who is preventing protection from themselves - but it also has a sustainable aspect as this system will be better suited in high speed areas like school zones, construction zones, accident zones, etc. SCSS is also related to intelligent transportation systems as well as systems within other countries that achieve vehicle safety and momentum safety for the safety of everyone else. The following sections of the proposal include the design of SCSS, the working system and the results of testing that prove working practicality.

## 2.Literature Review:
### *2.1 Review of Speed Control and Automatic Braking System*

The objective of this paper is to investigate the integrated potential of such a system to avoid collision and to prevent collision through the creation and design of these systems. The main electronic parts used for the creation of this system are ultrasonic sensors for obstacle detection and IR sensors for automatic braking. Essentially, if the vehicle is going fast enough; either the human driver doesn't apply brakes or he/she/they cannot apply them fast enough to prevent a crash.

*2.2 Literature Survey of Various Speed Control Vehicles:*

This is a systematic extensive survey of various speed control vehicles from the most basic with microcontrollers to the more advanced with GSM, GPS and other sensors (alcohol sensor, RFID sensor etc). This is a systematic and comparative study of such inventions and efforts and tries to determine relative benefits of newer inventions for accident prevention and life saving efforts on what can be considered the most dangerous day to day human activity - driving. Therefore, such an extensive literature survey allows for appreciation of inventions and efforts that require more tinkering and application moving forward on an interdisciplinary effort even with more complicated devices in this field. Ultimately, this literature survey suggests that such devices are rarely if ever correctly utilized from the onset in a functional state and thereafter adjustments are needed over time for future dilemmas which create the most functional product.

*2.3 Vehicle Speed Detection and Control System Based on Speed Restriction Signboards/Speed Management Systems Implementation*

This is a journal article that studies the vehicle speed detection and control system based on speed restriction signboards in other words, a vehicle has a speed camera affixed/connected to a traffic sign and it receives real time data from the camera as to its speed and when it should slow down. Speed detection is critical for accident prevention and expedited traffic maneuvering but real time longitudinal study can get things like this piloted in an international city - like the one this study took place in - which is done via simulation with variable predictability. Therefore, real time longitudinal data collection like this is related to how other speed control systems can operate.

*2.4 Intelligent Speed Adaptation Systems*

This is a journal article about intelligent speed adaptation (ISA) systems which operate via GPS and digital mapping to convey in real time the speed limit as someone is traveling and automatically applies the necessary measures to accommodate those speed limits (i.e. brakes too fast; accelerate too slow). This journal article seeks to determine the types of ISA systems (advisory; voluntary implementation; mandatory implementation) and what bad speed limits/understanding of regulations might warrant certain responses. There is a semi-structured approach taken relative to assessments and viability of ISA systems but a generalized limited implementation throughout is assessed.

2.5 *Devices and systems indicate speed limit.*
 *2.5.1. Speed Limit Warning Systems*
These systems alert the driver when they exceed a predefined speed.

    a. Speed Limit Indicator (SLI)

- Uses GPS and a database of speed limits.
- Displays the current speed limit on the dashboard or a heads-up display (HUD).
- Some vehicles have built-in SLI in their infotainment systems.

  b. Speed Warning System

- Allows drivers to set a speed threshold.
- Provides a visual and/or audio warning when the vehicle exceeds the set speed.

 *2.5.2. Intelligent Speed Assistance (ISA)*

A more advanced system that combines multiple technologies:

- Speed limits are detected using GPS data and road sign recognition.
- Alerts the driver when exceeding the limit.
- Some ISA systems automatically reduce engine power to prevent speeding.
- Mandated in new vehicles in the EU from 2022.

### 2.5.3. Heads-Up Display (HUD)

- Projects speed and speed limit warnings onto the windshield.
- Reduces distraction by keeping the driver's focus on the road.

### 2.5.4. Smartphone Apps

- Apps like Waze, Google Maps, and Speedometer GPS provide real-time speed warnings.
- Some apps use GPS and databases to warn about speed cameras and speed limits.

### 2.5.5. OBD-II Speed Alert Devices

- Plugged into the vehicle's OBD-II port.
- Monitors speed and provide alerts via sound or smartphone notifications.
- Common in fleet vehicles for tracking and safety.

### 2.5.6. Dash Cams with Speed Monitoring

- Some advanced dash cams have built-in GPS and speed monitoring.
- They display and warn the driver if speeding occurs.

### 2.5.7. External Speed Limit Signs & Roadside Devices

- Radar speed signs detect vehicle speed and display warnings.

Found in school zones and high-risk areas

## 3. Methodology

### 3.1 Problem Analysis & Research:

Over speeding causes about 30% of fatal accidents globally, as traditional enforcement methods fail to prevent real-time violations. Speed cameras, traffic signals, and manual policing rely on driver compliance, making them ineffective. An intelligent Speed Control Security System (SCSS) can autonomously regulate speed, enhancing road safety through real-time monitoring and intervention.

### 3.2 Component Acquiring:

Procured necessary hardware components: Arduino UNO, ESP-32 micro-Prosser, ESP-32 Camera Module, Hall Effect Sensor (RPM Sensor), Relay Module, 12-volt DC Motor. We also have manufactured several parts like the Compression Chamber, The Piston, Pinion, Brake Calliper, Calliper Mounts, Piston Rack

### 3.3 Electronic Report:

1. esp32 can operate in high temperature areas and in industrial use, its rating is between -40° C to +125° C it also consumes very less power as it uses multiple proprietary software's

2. its operating voltage is between 4.5V to 24V

3. the Esp32 is also compact in size as it only takes up space of 18.5mm x 15mm

4. it has a current rating of 260mA maximum (160~260mA)

### 3.4 Cad Report:

1. Calliper Model:

This is the model of a Fixed brake caliper which is connected to the knuckle of the car and creates friction between the rotor and brake pads to stop the vehicle

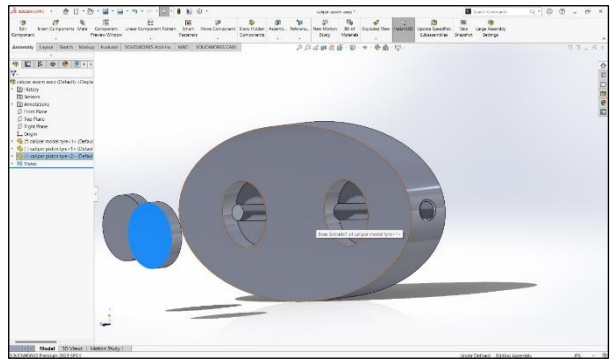

*Figure 1: 3D Calliper Model in Solid works.*

2. Compression chamber with piston and rack and pinion assembly

This assembly consists of several parts like the compression chamber, the piston with rack attached to it and the pinion

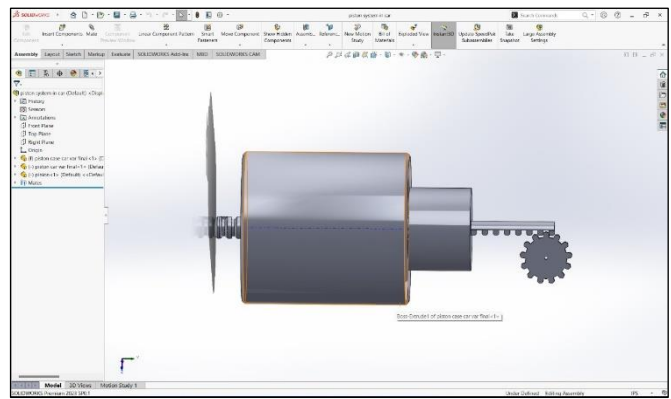

*Figure 2: Compression chamber with piston and rack and pinion assembly.*

1. Compression chamber

The work of the compression chamber is to hold the brake fluid and helps create pressure in the brake hose

2. Piston with Rack

This consists the piston which is mainly responsible to create the pressure in the compression chamber with the help of liner motion against the brake fluid which is created by the rack attached behind it

3. Pinion

The is used to convert the rotational movement of the DC motor into liner motion with the help of the Rack it also increases the total torque applied by the motor

## 3.5 Components and relevance

### 3.5.1 Arduino Uno:

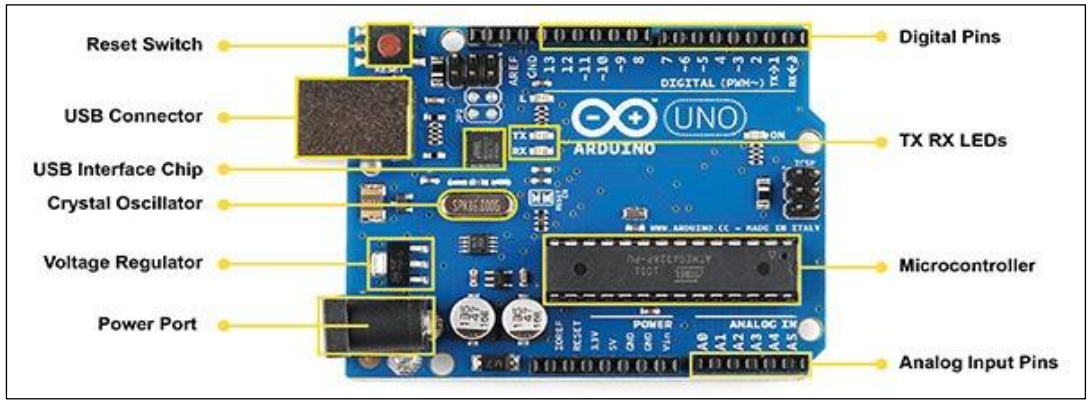

*Figure 3 : Arduino Uno*

The Arduino Uno is the most popular and extensively used microcontroller board; it contains the ATmega328P. This board is widely used in electronics because it is inexpensive, easy to program, and can be applied with an enormous array of sensors and other components.

In this project, the controller is

Arduino Uno; We have used it to process the data given to it by the ESP-32 micro processer which is then compiled by the Arduino UNO and it gives the signal of ON and OFF to the DC Motor

### 3.5.2. ESP-32 Micro Processor

The ESP32 is a strong System on Chip (SoC) microcontroller that includes Wi-Fi (802.11 b/g/n), dual mode Bluetooth version 4.2, and many other useful features. It is a better version of the older ESP8266 chip, with two separate processors that can run up to 240 MHz. Compared to the previous model, it has more GPIO pins—going from 17 to 36—and also has more PWM channels (up to 16). It also has 4MB of flash memory.

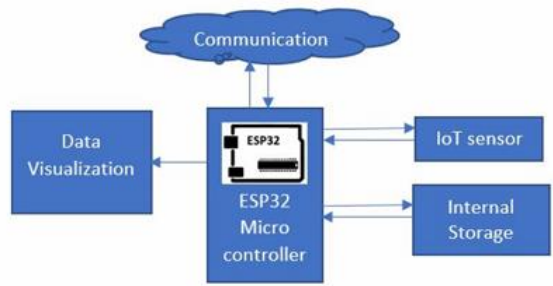

*Figure 4 : ESP-32 Workflow.*

### 3.5.3. DC Motor

A high-torque geared 12V DC motor consists of a DC motor and an integrated gearbox that reduces speed while increasing torque. When 12V DC is applied, the motor generates rotational motion through electromagnetic induction, where current flowing through the armature interacts with the stator's magnetic field to produce torque. The attached gearbox, using spur, planetary, or worm gears, reduces RPM and amplifies torque based on the gear ratio (e.g., 100:1 reduces speed 100 times but increases torque 100-fold). The motor's speed and direction can be controlled using PWM signals and H-Bridge circuits, making it ideal for robotics, automation, and industrial applications.

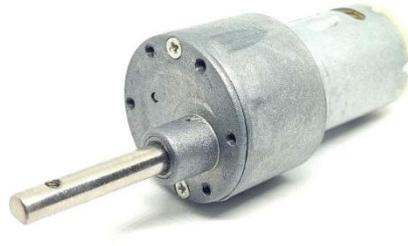

*Figure 5:  DC Motor used for actuation.*

### 3.5.4 ESP-32 Camera Module

Cameras used for IoT research, especially the ESP32-CAM, are a great fit for research projects. The key reasons are:

1. Easy to program: The ESP32-CAM works well with Arduino and has many ready-made coding examples.
2. Clock synchronization: It can sync its time with the internet using NTP.
3. Good storage & processing: It has enough memory and uses the OV2640 camera's built-in JPEG processing to reduce load on the processor.
4. Mounting options: There are many 3D-printed designs available for mounting and alignment.
5. Live view support: Although not built-in, projects like KISSE have developed libraries to enable it.
6. Affordable & replaceable: ESP32-CAM is low-cost, so having extra backups is easy

Here ESP32-CAM module is used in an IoT-based security alarm system. The ESP32-CAM is a low-cost microcontroller with WIFI and Bluetooth capabilities, supporting a camera module (OV2640) for image capture and processing. The prototype integrates a motion sensor (HC-SR501) and communicates through a Telegram bot, allowing users to receive security alerts and capture images remotely.

### 3.5.5. Arduino IDE

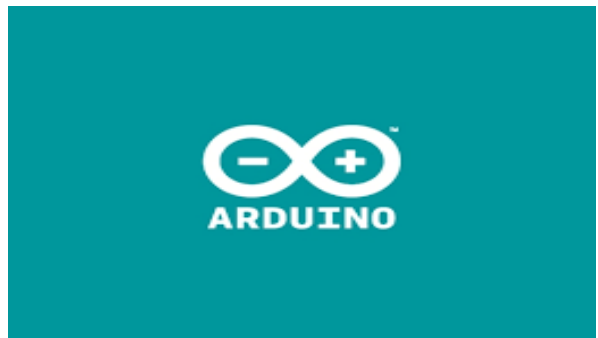

*Figure 6: Arduino IDE used for coding.*

The Arduino Integrated Development Environment (IDE) was essential in designing and programming the automatic irrigation system**.** It enabled the writing, uploading, and debugging of the code that governs the system's operations. Below is a detailed overview of how the **Arduino IDE** contributed to this project:

Use of Arduino IDE in the Automatic Irrigation System

### 3.5.5.1 Writing the Code

The Arduino IDE was utilized to develop the control logic for the system. The code was designed to calculate the data of the current speed of the car also to calculate the speed limit via the camera module and the compile this data and give output as an on and off signal to the DC

### 3.5.5.2. Uploading the Code

After writing the code, it was uploaded to the Arduino microcontroller (e.g., Arduino Uno) through the Arduino IDE. The steps involved were:

1. Connecting the Arduino board to a computer using a USB cable.
2. Selecting the appropriate board model and communication port in the Arduino IDE.
3. Clicking the Upload button to transfer the program to the microcontroller.

### 3.5.5.3 Debugging and Testing

The Arduino IDE has a feature called the Serial Monitor which was useful in the process of debugging and testing. It allowed real-time monitoring of sensor readings and system status to ensure proper functionality.

By utilizing the Arduino IDE, the development, programming, and troubleshooting of the speed control system became efficient and precise

## 4. Results and Discussion

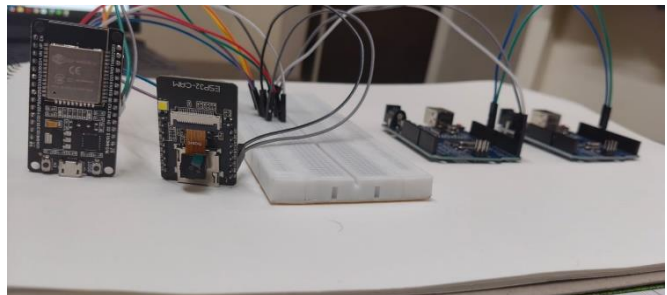

*Figure 7: Entire electronic circuit involved in this project.*

How the system works
### 4.1 Cam Module:
The cam module captures the number of lanes available on the road the vehicle is currently driving on which is used to find out the speed limit of the particular road
### 4.2 Hall Effect Sensor:
The hall effect sensor works as an RPM sensor of the car witch accurately measures the current speed of the vehicle
### 4.3 ESP-32
The ESP-32 is acts like the mother board of the circuit it assigns the speed limits to the data of, amount of lanes given by the cam module it them compiles this with the data output given by the hall effect sensor to find out the threshold speed i.e. Legal speed limit of the road this compiled data is further supplied to an Arduino UNO
### 4.4 Arduino UNO

The Arduino UNO takes the data from the ESP-32 module to give the output to the DC motor, this is an fluctuating on and off output so that the brakes do not lock and cause an accident.

5. Future Improvements:

*5.1. Ai-powered dynamic control*: SCSS could be enhanced with ai to continuously assess road conditions and patterns as well as vehicle/human responses in proximity for an even more naturalized speed.

*5.2. Connected vehicle technology*: SCSS could be connected to intelligent traffic systems, GPS and continuous road quality assessments for greater communicative potential.

*5.3. Emergency override*: SCSS could possess an emergency manual override that wouldn't down the entire system should a human driver have to take the wheel in an emergency.

*5.4. Advanced sensors*: SCSS could feature enhanced lidar, radar and computer vision readings for greater trust without outside confirmation of speed limit or project intervention.

*5.5. Lower cost and universal accessibility*: SCSS could be a more affordable and universally accessible option for all vehicles as costs decline and integration becomes more universal amongst systems.

*5.6. Legal and regulatory policy backing*: SCSS could possess legal and regulatory backing for international connected vehicle policy support for speed limiting devices.

5.7. In weather imperfections: SCSS could be effective in bad weather - meaning heavy rain, mist/fog or snow - which would impair speed detection or limit detection of surroundings.

5.8. Vehicle to vehicle connectivity SCSS could communicate with other vehicles with such tech for an even more collaborative speed determination.

6. Conclusion

The Speed Control Security System (SCSS) revolutionizes global over speeding situations which is the leading cause of vehicular accidents and injuries/fatalities. Where typical over speeding enforcement occurs via detection, punishment, a level of driver willingness, the SCSS which intends to learn speed passively as a prevention for further accidents, is a revolutionary vehicle technology for safety and situational awareness in real time for what the vehicle needs to do. The SCSS is an efficient, revolutionary speed control solution that intends to rapidly assess sensor technologies, microcontroller intercommunication and wireless communication devices for automatically speed based intercession during critical situations like school zones, work zones, and pedestrian crosswalks or low speed situations on interstate highways which are highly accident prone.

While the benefits of the SCSS as a revolutionary, lifesaving, advanced feature are great, limitations for effective implementation include costs, operating feasibility, legal challenges and potential social opposition. However, with the onset of artificial intelligence systems, vehicle-to-infrastructure (V2I) connectivity and increased integration with adaptive speed control systems, not only does the SCSS become more effective and efficient over time, but other developments support the reliability of the SCSS through real time assessments. Assessments can be made for determinations for SCSS functioning for sensor assessments and emergency response premption for SCSS speed determinations that may conflict with the desires for increased speed by humans.

As the future transitions into intelligent transport systems and self driving vehicle worlds, the SCSS can be suggested as a standard across all vehicles as an innate operational feature which transforms the idea of safety and preventative measures of accident prevention. Therefore, it is up to municipalities and lawmakers to align with automotive manufacturers to create standardized regulations and adjusted costs to adapt the SCSS into all new vehicles going forward because the extensive literature review supports accident mitigating/life saving probability that will create safer, more intelligent transport systems for all.

7. Refrences: